\documentclass[preprintnumbers,a4paper,twocolumn,floats,aps,nofootinbib]{revtex4}
\usepackage{graphicx}
\usepackage{amsmath,amssymb,amsfonts}
\usepackage[dvips]{hyperref}
\begin{document}
\title{ DAMA and CoGeNT without astrophysical uncertainties}

\preprint{OUTP-11-43-P}

\author{Christopher McCabe}
\email{c.mccabe1@physics.ox.ac.uk}

\affiliation{Rudolf Peierls Centre for Theoretical Physics, University of Oxford, Oxford OX1 3NP, United Kingdom}

\begin{abstract} 
The CoGeNT collaboration has reported evidence of an annual modulation in its first fifteen months of data. Here we compare the amplitude and phase of this signal to the modulation observed by the DAMA collaboration, assuming that both arise due to elastically scattering dark matter (DM). We directly map the CoGeNT signal to the DAMA detector without specifying any astrophysical parameters and compare this with the signal measured by DAMA. We also compare with constraints from CDMS II and XENON10. We find that DM of mass 5-14 GeV that couples equally to protons and neutrons is strongly disfavoured. Isospin-violating DM fares better but requires a boosted modulation fraction.
 \end{abstract} 

\date{\today}
\maketitle

\section{Introduction} 

There is strong observational evidence for a large abundance of particle dark matter (DM) in our Universe. Detecting this matter through non-gravitational interactions is a great challenge. A potentially characteristic signal arising from particle DM, however, is the presence of an annual modulation in the event rate at direct detection experiments \cite{Drukier:1986tm, Freese:1987wu}. These experiments  aim to detect the energy deposited by a nucleus after scattering with DM. The modulation signal arises due to the motion of the Earth relative to the galactic halo and peaks when the Earth is travelling fastest with respect to the halo. For most dark matter halo models, this occurs in late May or early June (see e.g.\ \cite{Fornengo:2003fm,Green:2003yh}).

Two experiments have now detected an annual modulation signal that has many features consistent with a signal arising from DM. The DAMA collaboration \cite{Bernabei:2008yi, Bernabei:2010mq} has observed an annual modulation over thirteen years; first with the DAMA/NaI setup and later with the DAMA/LIBRA upgrade. The large mass of the target material and the long exposure time mean that the statistical nature of the modulation is beyond question. More recently, the CoGeNT collaboration \cite{Aalseth:2011wp} has presented tentative evidence for an annual modulation after analysing fifteen months of data.

In this paper, we assume both signals arise due to DM and study the consistency of the two signals assuming the DM scatters elastically. In particular, we consider whether the phase and amplitude of the modulations are consistent. Previous studies \cite{Frandsen:2011ts, Hooper:2011hd, Foot:2011pi, Belli:2011kw, Schwetz:2011xm} have shown that DAMA and CoGeNT can be in agreement for DM of mass $\sim5$-$14$ GeV that scatters elastically via a spin-independent interaction, albeit with tension from the CDMS \cite{Ahmed:2010wy}, XENON10 \cite{Angle:2011th}, XENON100 \cite{Aprile:2011hi} and SIMPLE \cite{Felizardo:2011uw} experiments.\footnote{See \cite{Collar:2011kf, Collar:2011wq, Collar:2011kr} for a critical discussion of these experiments.}  However, it is well known that interpreting the signals at direct detection experiments is sensitive to many uncertainties, particularly uncertainties in astrophysical parameters (see e.g.\ \cite{Fairbairn:2008gz, MarchRussell:2008dy, Kuhlen:2009vh, McCabe:2010zh, Green:2010gw, Lisanti:2010qx, Arina:2011si}). Therefore, we follow the approach of \cite{Fox:2010bz}, which specifies how to directly map experimental signals from one detector to another, allowing a comparison without specifying any astrophysical parameters. Given the modulation observed at CoGeNT, we calculate the peak day and modulation amplitude expected at DAMA, which we compare with that observed at DAMA. The phase and amplitude will be the same if both modulations arise from DM. We finish by considering constraints from the low threshold analysis of CDMS II \cite{Ahmed:2010wy, Akerib:2010pv} and XENON10 \cite{Angle:2011th}.

\section{Dark matter detection}
\label{sec:overview}

We briefly review the formalism behind direct detection experiments and define our notation. The differential event rate for elastic spin-independent DM-nucleus scattering in counts per day per unit nucleus mass per unit exposure time per unit energy (cpd/kg/keV) as a function of recoil energy $E_R$ is
\begin{equation}
\frac{dR}{dE_R}=\frac{\rho_{\chi} \sigma_n}{2 m_{\chi} \mu^2_{n \chi}} \frac{C_{T}}{f_n^2}  F^2(E_R) \epsilon (E_R) g(v_{\text{min}}, t) .
\end{equation}
Here $m_\chi$ is the DM mass, $\rho_\chi$ the local DM density, $\mu_{n \chi}$ the DM-nucleon reduced mass and $\sigma_n$ is the DM-neutron cross-section at zero momentum transfer in the elastic limit. $F(E_R)$ is the form factor which we take as
\begin{equation}
F^2(E_R)=\left(\frac{3 j_1(qR)}{qR}\right)^2 e^{-q^2 s^2} ,
\end{equation}
with $q=\sqrt{2 m_N E_R}$, $R=\sqrt{c^2+\frac{7}{3} \pi^2 a^2 -5 s^2}$, \mbox{$c=1.23 A^{1/3}-0.60$} fm, $s=0.9$ fm, $a=0.52$ fm and $m_N$ is the mass of the target nucleus. We define $C_{T}$ as
\begin{equation}
C_T=\kappa (f_p Z +f_n (A-Z))^2 ,
\end{equation}
where $\kappa$ is the detector mass fraction of the target nucleus, A and Z are the nucleon and proton number respectively, while $f_n$ and $f_p$ encode the coupling of the DM to neutrons and protons respectively. $\epsilon(E_R)$ is the efficiency of the detector and, in general, depends on $E_R$.

Finally, $g(v_{\text{min}}, t)$ encodes all the information about the DM velocity distribution:
\begin{equation}
g(v_{\text{min}}, t)=\int^\infty_{v_{\text{min}}}  \frac{f_{\text{local}}(\vec{v},t)}{v} d^3v  ,
\end{equation}
where $f_{\text{local}}$ is the local DM velocity distribution. $v_{\text{min}}$ is the minimum DM speed required for a nucleus to recoil with energy $E_R$:
\begin{equation}\label{vmin}
v_{\text{min}}=\sqrt{\frac{m_N E_R}{2 \mu^2}} ,
\end{equation}
where $\mu$ is the reduced mass of the DM-nucleus system.

\section{A formalism free from astrophysics}
\label{sec:astrofree}

Following the approach of \cite{Fox:2010bz}, we wish to map the differential event rate into $v_{\text{min}}$ space. As we will show, when the $v_{\text{min}}$ space probed by two different experiments is the same, we can compare them directly without making any assumptions about any astrophysical parameters. This allows us to map the peak day and amplitude of the CoGeNT modulation onto the DAMA detector, which can be compared with the signal that DAMA observes. Below we briefly recap the relevant theory from \cite{Fox:2010bz}.

Using Eqn.\ \ref{vmin} we rewrite the event rate in $v_{\text{min}}$ space:
\begin{eqnarray}
\label{Ratevspace2} R(t)&=\frac{2 \rho_{\chi} \sigma_n}{m_N m_\chi}\frac{\mu^2}{\mu_{n \chi}^2}\frac{C_{T}}{f_n^2} \int^{v_{\text{high}}}_{v_{\text{low}}}F^2(E_R) \epsilon (E_R) v g(v, t) dv \\
& \approx \frac{2 \rho_{\chi} \sigma_n}{m_N m_\chi}\frac{\mu^2}{\mu_{n \chi}^2}\frac{C_{T}}{f_n^2} \bar{F}^2(E_R) \bar{\epsilon} (E_R) \int^{v_{\text{high}}}_{v_{\text{low}}}  v g(v, t) dv \label{Ratevspace}
\end{eqnarray}
where $E_R=2 \mu^2 v_{\text{min}}/m_N$ is now a function in $v_{\text{min}}$ space. $\epsilon(E_R)$ and $F(E_R)$ are almost flat across the range of recoil energies we consider at DAMA, CoGeNT and later CDMS II. Therefore, we remove them from the integral and evaluate them at the average value across the domain of integration. The error introduced by this approximation is less than a few percent.

All of the physics that determines the peak day of the modulation is encapsulated in the integral $\int^{v_{\text{high}}}_{v_{\text{low}}}  v g(v, t) dv$. By arranging to have the same integration limits for DAMA and CoGeNT, the respective modulations will have the same peak day if they both arise from DM scattering. Given an energy range at DAMA (D), we can invert Eqn.\ \ref{vmin} to find the energy range at CoGeNT (C) which spans the same $v_{\text{min}}$ space. Doing so we find:
\begin{equation}\label{Econvert}
[E_{\text{low}}^{C}, E_{\text{high}}^{C}]=\frac{\mu_{C}^2 m_N^{D}}{\mu_D^2 m_N^{C}} [E_{\text{low}}^{D}, E_{\text{high}}^{D}] .
\end{equation}

Given the modulation amplitude $\Delta R=(R(t_{\text{max}})-R(t_{\text{min}}))/2$ observed by CoGeNT (in cpd/kg), and a specific choice of $m_{\chi}$ and $C_T$, we can invert Eqn.\ \ref{Ratevspace} to solve for $\int^{v_{\text{high}}}_{v_{\text{low}}}  v g(v, t) dv$. This allows us to calculate the amplitude expected at DAMA (in cpd/kg):
\begin{equation}\label{amplitude}
\Delta R^D_{\text{expec}} = \frac{ \bar{\epsilon}_D(E^D_R) \bar{F}^2_D(E^D_R)}{\bar{\epsilon}_C (E^C_R) \bar{F}^2_C(E^C_R)} \frac{C_T^{D}}{C_T^{C}} \frac{m_N^{C} \mu_D^2}{m_N^{D} \mu_C^2} \Delta R^C_{\text{obs}}
\end{equation}
We compare this with what is observed at DAMA. If both modulations arise from DM scattering, the expected and observed amplitude will be the same. Since no astrophysical parameters enter into this formula, the two experiments can be compared free from astrophysical uncertainties. 

\section{Comparing DAMA and CoGeNT}

\begin{figure}[!t]
{
\includegraphics[width=0.99\columnwidth]{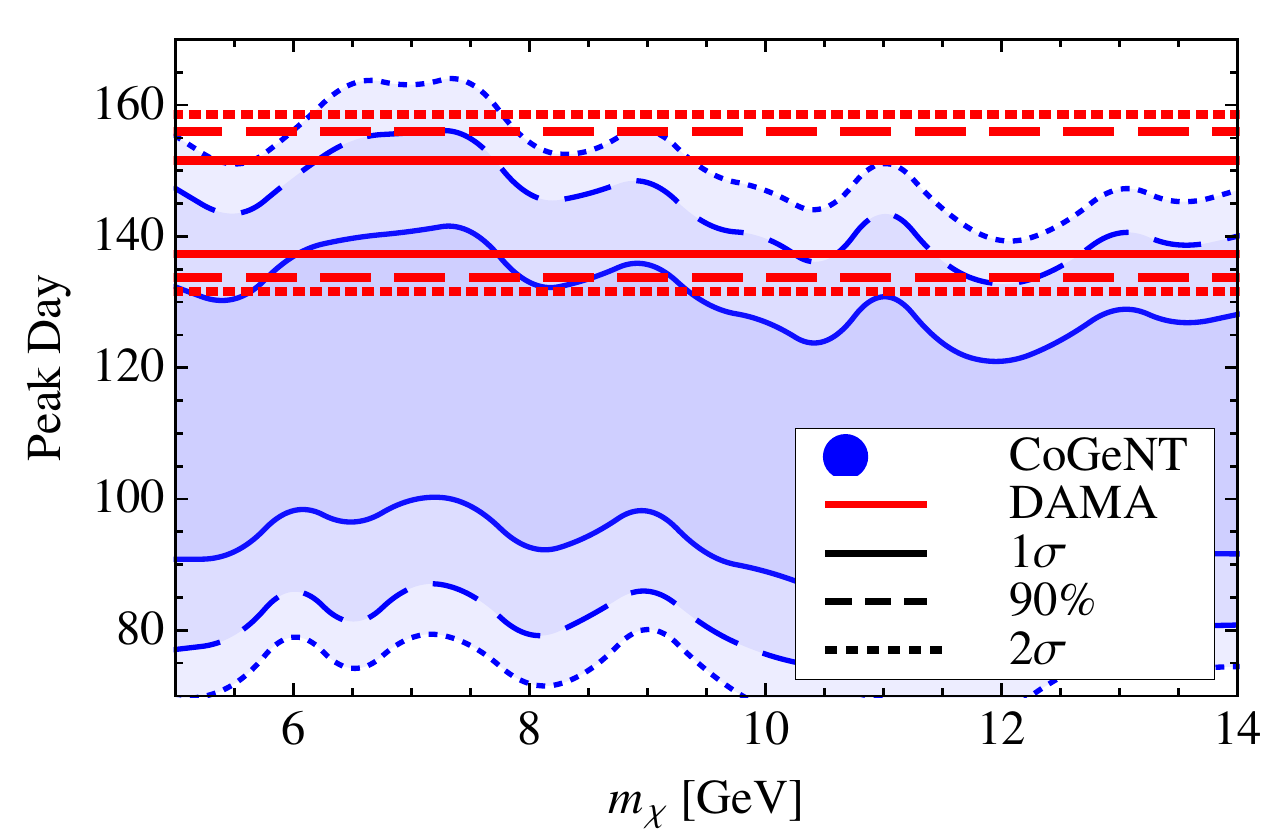}
}
\caption{The peak day of the modulation signals measured by DAMA (red) and CoGeNT (blue) for different values of $m_{\chi}$. At DAMA, we fit to the data in the 2-6 keVee energy range measured over thirteen annual cycles. At CoGeNT, we fit to the data in the energy range determined using Eqn.\ \ref{Econvert}. It is clear that the CoGeNT modulation generally peaks earlier in the year than DAMA's modulation signal, although there is agreement at $1\sigma$ for $m_{\chi}\sim7$ GeV. The Standard Halo Model predicts the modulation peaks on Day 152 = 2nd June.}
\label{fig0}
\end{figure}

\begin{figure*}[!th]
{
\includegraphics[width=0.88\columnwidth]{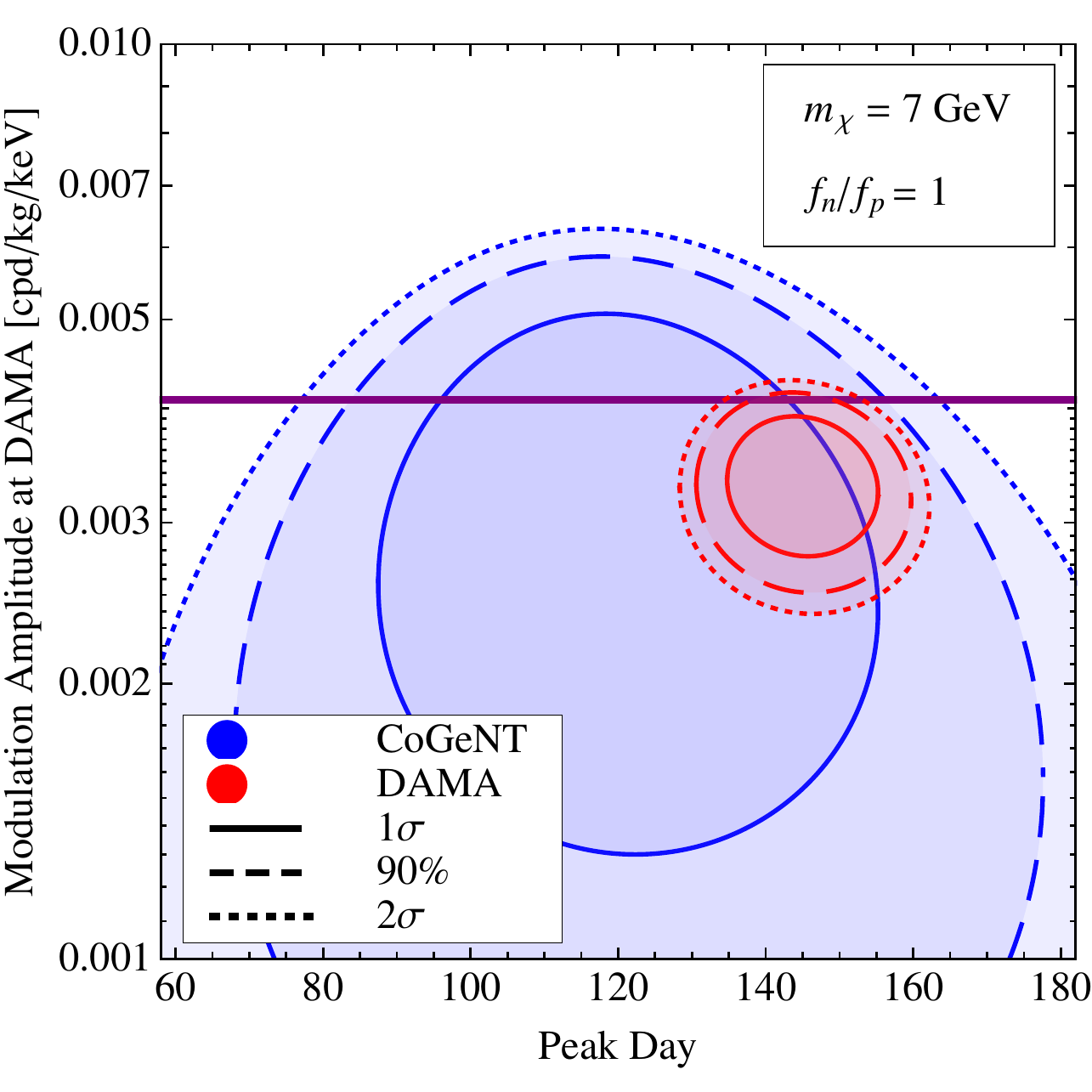}
\includegraphics[width=0.88\columnwidth]{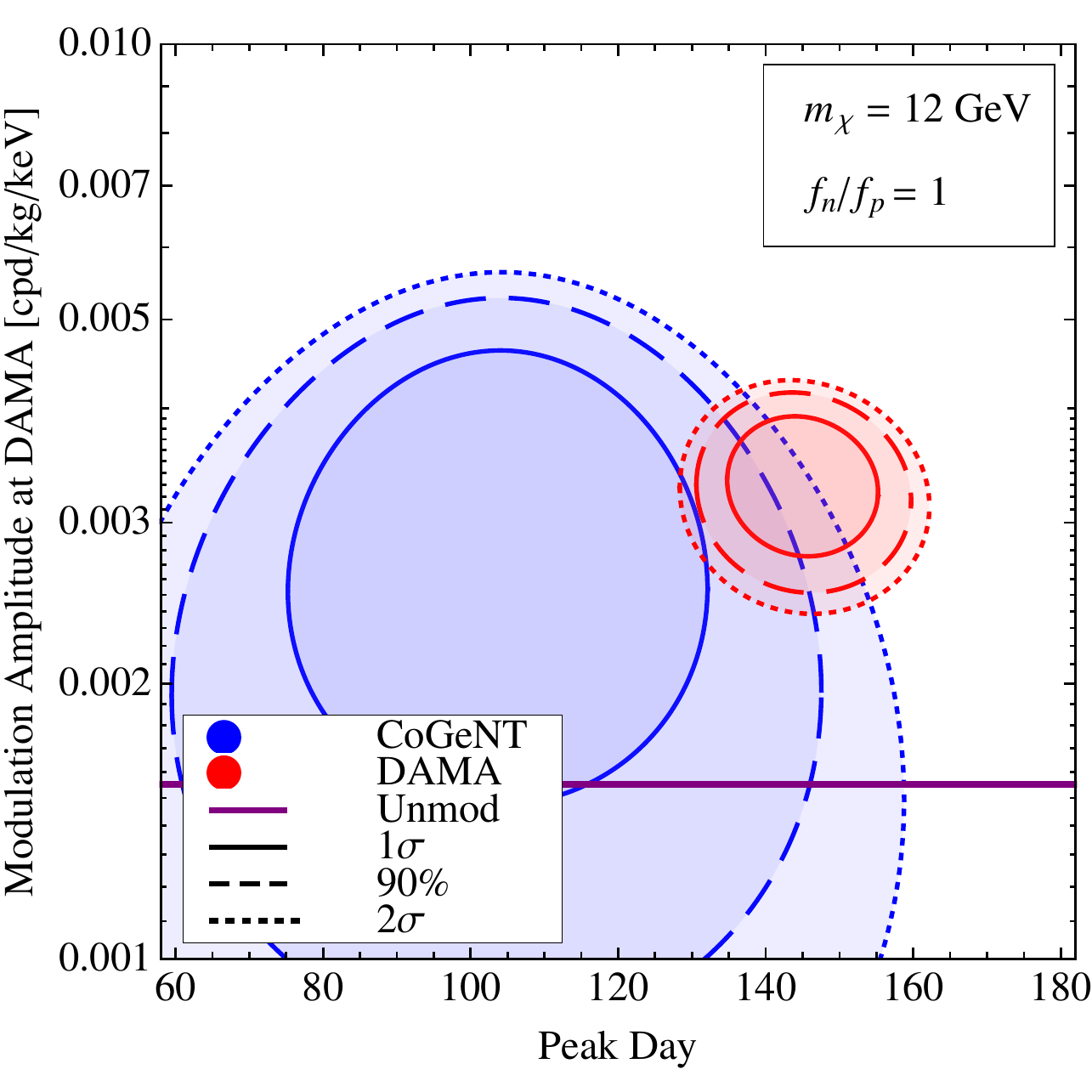}
\includegraphics[width=0.88\columnwidth]{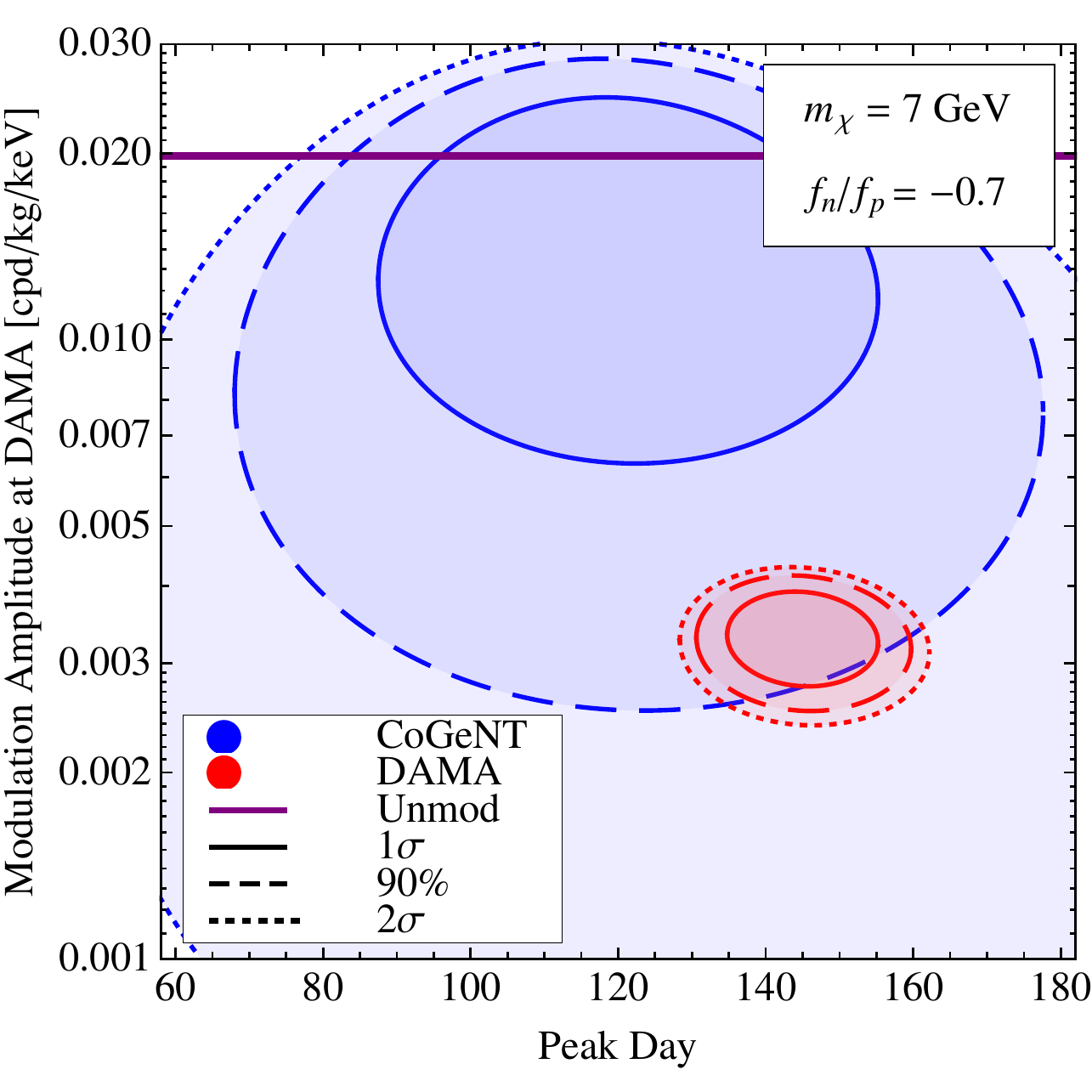}
\includegraphics[width=0.88\columnwidth]{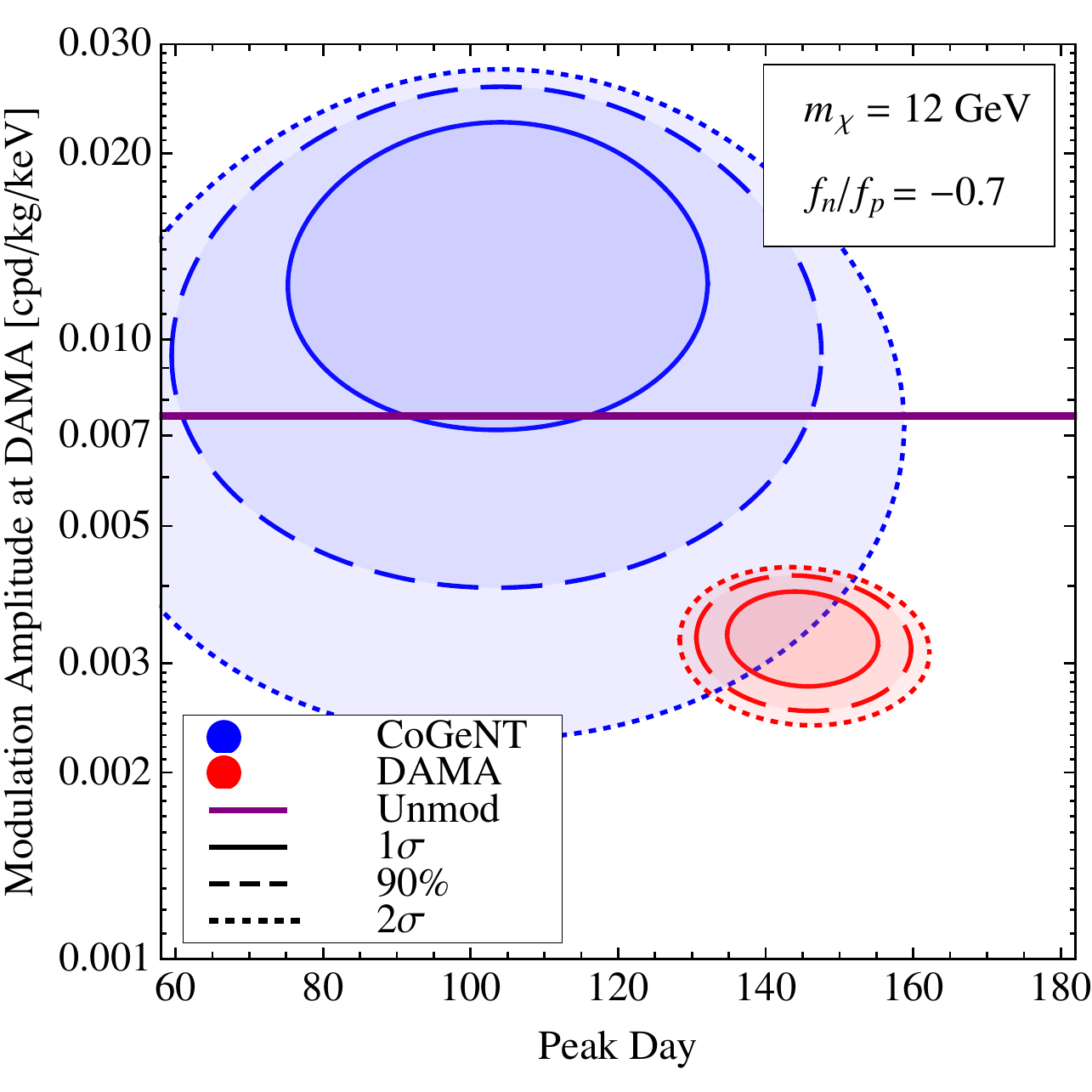}
}
\caption{Best fit regions for DAMA and CoGeNT from a two parameter fit to the amplitude and peak day of the respective modulation signals. Day 152 = 2nd June. The blue region shows the amplitude and peak day expected at the DAMA detector based on what CoGeNT observe. The red contours show the regions obtained from fitting to the 2-6 keVee DAMA data in \cite{Bernabei:2008yi, Bernabei:2010mq}. In the left (right) panels $m_{\chi} =7\, (12)$ GeV. In the upper (lower) panels $f_n/f_p=1\,(-0.7)$. The purple solid line indicates the unmodulated rate at CoGeMT. Modulation amplitudes above this line are excluded. Note the lower panels have a different scale for the modulation amplitude.}
\label{fig1}
\end{figure*}

For the purposes of this analysis, we assume the DM only scatters off the sodium nuclei at DAMA. This is a very good approximation for the light DM that we consider. DAMA presents their time-dependent results in fixed energy ranges: 2-4 keVee, 2-5 keVee and 2-6 keVee. This can be converted to keV by dividing by the quenching factor for sodium, which we take as $q_{\text{Na}}=0.3$. Varying $q_{\text{Na}}$ does not lead to significant differences in our results. We do not include channelling, as theoretical calculations indicate this is small \cite{Bozorgnia:2010xy}.

The CoGeNT collaboration has released the time-stamped data for independent analysis. We bin their data in the energy range that  spans the same $v_{\text{min}}$ space as DAMA. Statistics at CoGeNT are limited, therefore we concentrate on the 2-6 keVee bin at DAMA, as this corresponds to the largest energy range at CoGeNT and will therefore contain the most events. To convert from keV to keVee at CoGeNT, we use $E[\text{keVee}]=0.199 E[\text{keV}]^{1.12}$ \cite{Collaremail}. For each energy range, we divide the 458 days of data into 15 time bins with a width of 30 days, and one final bin with a width of 8 days, correcting, when appropriate, for periods when the detector was not running.

In Fig.\ \ref{fig0} we show the peak day found at DAMA and CoGeNT. The regions between the red lines indicate the preferred regions at $1\sigma$ (solid), $90\%$ (dashed) and $2\sigma$ (dotted) after fitting to the residual events as a function of time measured by DAMA/NaI \cite{Bernabei:2008yi} and DAMA/LIBRA \cite{Bernabei:2010mq} over thirteen annual cycles. For each mass, we bin the data at CoGeNT in the relevant range determined from Eqn.\ \ref{Econvert} and plot the results in blue. For both DAMA and CoGeNT, we fix the period of the modulation to be one year and consider variations of the phase that satisfy $\chi^2\leq \chi^2_{\text{min}}+\Delta \chi^2$. The variation in the peak day for different values of $m_{\chi}$ is expected, since for each mass the data is binned in a different energy range. As well as purely statistical fluctuations in the observed count rate, N-body simulations have shown that variations in the peak day as a function of energy should be expected \cite{Kuhlen:2009vh} and are also observed by DAMA (see Fig.\ 9 of \cite{Bernabei:2010mq}). Generically, the CoGeNT modulation peaks earlier in the year than DAMA. It is only for $m_{\chi}\sim7$ GeV that there is agreement at $1\sigma$.

Fig.\ \ref{fig1} shows the $1\sigma$ (solid), $90\%$ (dashed) and $2\sigma$ (dotted) contours in the peak day against modulation amplitude plane for $m_{\chi}=7$ GeV (left panels) and $m_{\chi}=12$ GeV (right panels). In the upper panels, we choose $f_n/f_p=1$ while in the lower panels $f_n/f_p=-0.7$. The choice $f_n/f_p=-0.7$ is phenomenologically motivated by the desire to suppress the event rate at xenon experiments \cite{Giuliani:2005my, Chang:2010yk, Feng:2011vu,  Frandsen:2011ts, DelNobile:2011je}. Given the modulation amplitude measured at CoGeNT, we show in blue the preferred region calculated using Eqn.\ \ref{amplitude} for the expected amplitude at DAMA. The red region show the fit to the DAMA data measured over thirteen cycles. In our fits, we fix the period to be one year and subtract a constant rate at CoGeNT, determined from the value which minimises the $\chi^2$. We proceed to find the values of the modulation amplitude and peak day which satisfy $\chi^2\leq \chi^2_{\text{min}}+\Delta \chi^2$.

The energy ranges at CoGeNT for $m_{\chi}=7$ GeV and $m_{\chi}=12$ GeV, determined from Eqn.\ \ref{Econvert}, are 0.70 -- 2.38 keVee and 0.87 -- 2.96 keVee respectively.  In these energy ranges, the presence of an annual modulation is preferred over a constant event rate at $2.0 \sigma$ and $2.4 \sigma$ respectively. The purple solid horizontal line indicates the constraint from the unmodulated rate measured by CoGeNT. We integrate the unmodulated rate in the energy range, calculated using Eqn.\ \ref{Econvert}, after subtracting the L-shell EC contribution and a constant background. We map this onto the DAMA detector using Eqn.\  \ref{amplitude}. A modulation amplitude above this line predicts a modulated rate which is larger than the unmodulated rate. A modulation amplitude above this line is excluded since it would predict a negative event rate in the winter. Although the regions favoured by CoGeNT are large due to the low statistics, we can already see that $m_{\chi}\sim7$ GeV is preferred over 12 GeV and that the fit is worse for $f_n/f_p=-0.7$, where a larger modulation amplitude is generally predicted. 

For 7 GeV and $f_n/f_p=1$ (upper left panel), we see that there is good agreement between the DAMA and CoGeNT modulation amplitude and peak day. However, the modulation amplitude is close to the unmodulated limit, so a large modulation fraction ($\sim 70\%$) is required to be consistent. We will return to this issue in the next section. For $f_n/f_p=-0.7$ (lower left panel), the best fit modulation amplitude predicted by CoGeNT is much larger than that observed by DAMA. However, the $90\%$ regions do overlap and a smaller modulation fraction ($\sim 15\%$) is required to be consistent with the unmodulated rate.

For 12 GeV and $f_n/f_p=1$ (upper right panel), the CoGeNT unmodulated rate excludes all of the DAMA region. For $f_n/f_p=-0.7$ (lower right panel) it is only the DAMA and CoGeNT $2 \sigma$ regions that are in agreement. We thus conclude that a 12 GeV DM particle with $f_n/f_p=1$ or $-0.7$ is disfavoured, before applying constraints from other experiments.

\section{Other constraints}

\begin{figure}[!t]
{
\includegraphics[width=0.99\columnwidth]{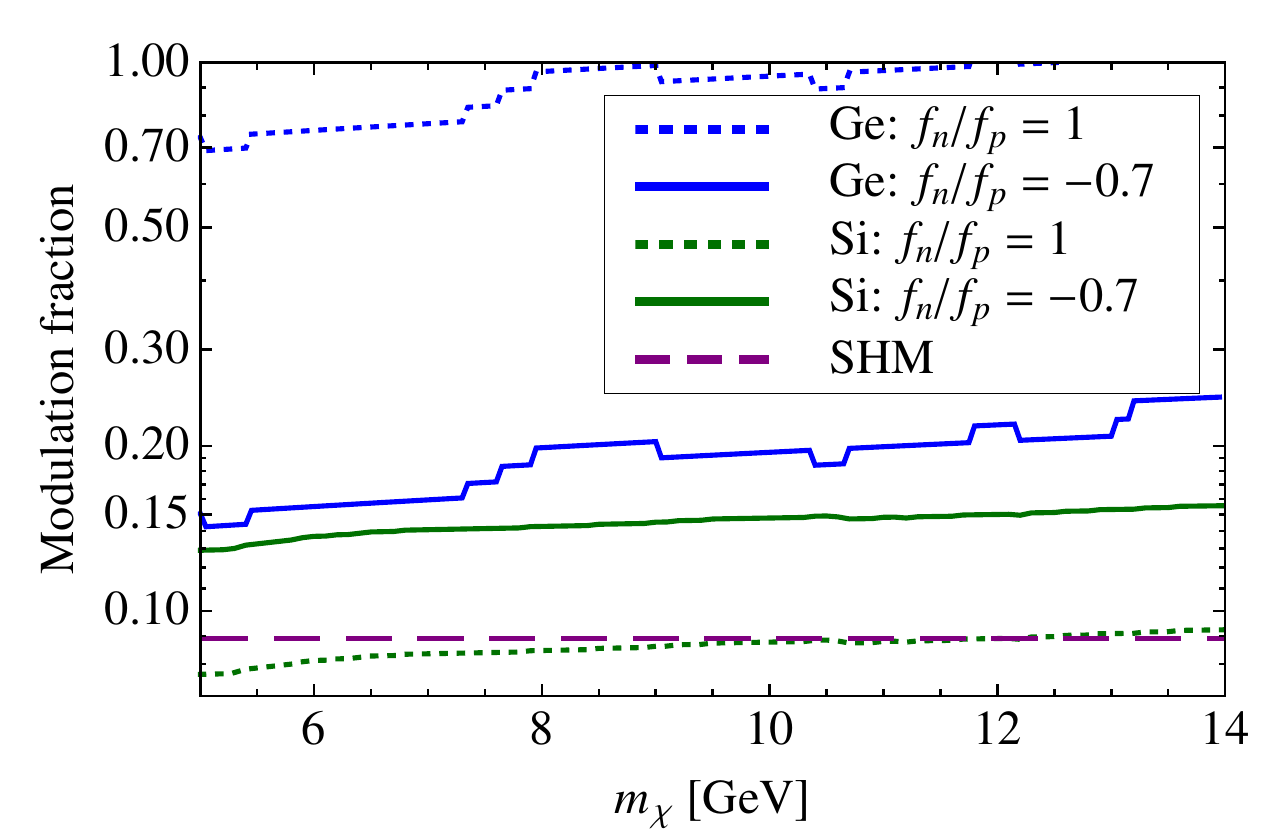}
}
\caption{The $90\%$ lower confidence limit on the fractional modulation required to be compatible with CDMS II. The blue (green) lines indicate constraints from the Ge (Si) analysis. The dotted (solid) line is for $f_n/f_p=1\, (-0.7)$. For comparison, the purple horizontal dashed line shows the modulation fraction from the Standard Halo Model (SHM).}
\label{figcdms}
\end{figure}

CDMS II has analysed data over a similar energy range as CoGeNT in its low threshold Soudan Underground Laboratory germanium analysis \cite{Ahmed:2010wy} and Stanford Underground Facility silicon analysis \cite{Akerib:2010pv}.\footnote{Silicon data has also been presented in \cite{Filippini:2008zz}. We do not consider it here due to uncertainties in calibrating the energy scale for nuclear recoils near threshold \cite{JPFilippini2011}.} Limits on an annual modulation signal have not been published but we can restrict the unmodulated rate, which we use to constrain the modulation fraction 
\begin{equation}
\frac{R(t_{\text{max}})-R(t_{\text{min}})}{R(t_{\text{max}})+R(t_{\text{min}})} .
\end{equation}

For the germanium analysis, we use the 35 kg-days of data collected between October 2006 and September 2008 from the T1Z5 detector, which has the best ionisation resolution. For the silicon analysis, we use the 24.64 kg-days collected between December 2001 and June 2002. To set conservative limits, we assume that all of the observed events arise due to DM. In Fig.\ \ref{figcdms}, the $90\%$ lower confidence limits on the modulation fraction for $f_n/f_p=1$ (dotted) and $f_n/f_p=-0.7$ (solid) are shown. The blue (green) lines are from the germanium (silicon) analysis. We assume the modulation amplitude at DAMA is 0.0028 cpd/kg/keV, which we see from Fig.\ \ref{fig1}, is on the lower edge of the CoGeNT $90\%$ region for $m_{\chi}=7$ GeV. For each value of the mass, we use Eqn.\ \ref{Econvert} to find the energy range at CDMS II which spans the same $v_{\text{min}}$ space as the 2-6 keVee range at DAMA. As the energy range varies for each mass, discrete jumps occur whenever a measured event at CDMS II enters or leaves the energy range. For comparison, the purple dashed horizontal line shows the modulation fraction predicted from the Standard Halo Model (SHM). We see that the modulation fraction required at DAMA and CoGeNT to be compatible with CDMS II-germanium is larger for both choices of $f_n/f_p$, while for CDMS II-silicon, is larger for $f_n/f_p=-0.7$.

The 2-6 keVee range at DAMA corresponds to $1.84$-5.52 keV at a xenon-target experiment (for $m_{\chi}$=7 GeV). We do not consider limits from XENON100 \cite{Aprile:2011hi}, as $\mathcal{L}_{\text{eff}}$ has not been measured below 3 keV \cite{Plante:2011hw}. However, we can apply the limits from the S2 only analysis of XENON10 \cite{Angle:2011th}, which has a low energy threshold of 1.4 keV. The 15 kg-days of data were collected between 23rd August and 14th September. Since this is approximately half way between the maximum and minimum of the CoGeNT and DAMA modulation signals, we assume the measured rate is the same as the unmodulated rate. We use the parameterization for $\mathcal{Q}_{\text{y}}$ given in \cite{Angle:2011th} and assume all events arise due to DM. We do not apply the edge (in z) event rejection. For each mass, we again use Eqn.\ \ref{Econvert} to find the energy range that spans the same $v_{\text{min}}$ space as the 2-6 keVee range at DAMA, and assume a modulation amplitude of 0.0028 cpd/kg/keV. For $f_n/f_p=1$, we find the required modulation fraction is $>100\%$ for all masses we consider. Hence, under the assumptions we have made, XENON10 excludes all parameter space. For $f_n/f_p=-0.7$ the constraints are severely weakened; a modulation fraction greater than $\sim2\%$ is required over the whole mass range, which is easily achieveable.

\section{Summary and Conclusions}

We have presented a comparison of the CoGeNT and DAMA modulation signals free from astrophysical uncertainties, having assumed that both modulation signals arise due to elastically scattering DM.

We found that the peak day of the CoGeNT modulation is always earlier than the peak day of the DAMA modulation. The $1 \sigma$ confidence regions do overlap in some parameter space, but the best fit points typically differ by $\sim 30$ days. The SHM, which assumes the DM is distributed isotropically, predicts a peak day close to 2nd June. If CoGeNT continue to measure a peak day at the lower edge of the DAMA confidence regions (around mid-May), there will be interesting consequences for DM galactic halo models.

For DM that couples equally to protons and neutrons ($f_n/f_p=1$), we found the measured modulation amplitude and peak day at DAMA is consistent with that expected based on the CoGeNT results. However, the XENON10 S2 analysis excludes the whole mass range in question. Even if we were to ignore the XENON10 analysis, tension remains with the unmodulated CoGeNT rate and the constraint from CDMS II-germanium, which require large modulation fractions. For $m_{\chi}=12$ GeV, the CoGeNT unmodulated rate excludes the DAMA modulation signal. Moreover, for all masses we consider, the modulation fraction needs to be larger than $\sim 70\%$ to be consistent with the low energy analysis of CDMS II-germanium. The SHM predicts $\sim 9\%$, which is significantly smaller. Such large deviations from the SHM seem unrealistic. Therefore, based on the constraints from XENON10, CDMS II and the CoGeNT unmodulated rate, we conclude that elastically scattering DM with $f_n/f_p=1$ is unlikely to be the source of the DAMA and CoGeNT modulation signals.

We also considered DM with isospin violating couplings $f_n/f_p=-0.7$. For this choice, the expected modulation amplitude at DAMA, calculated from the CoGeNT measurement, is generically higher than that observed at DAMA. For $m_{\chi}=12$ GeV, we find that only the $2 \sigma$ CoGeNT and DAMA regions overlap. However, for $m_{\chi}=7$ GeV, amplitudes at the lower end of the CoGeNT $90\%$ region are compatible. Furthermore, the CDMS II germanium and silicon constraints on the modulation fraction are milder, typically requiring modulation fractions $\sim 15 \%$. This is still larger than that from the SHM, so if the DAMA and CoGeNT signals do arise from DM, the constraints from CDMS II indicate that there will be interesting consequences for the DM galactic halo model.

It is clear that the evidence for a modulation in the CoGeNT data is still tentative and that much more data is required to definitively confirm a modulation. Fortunately, the CoGeNT-4 upgrade should provide much more data and significantly shrink the CoGeNT best fit regions. With more data, the approach presented here will serve as a useful complementary test on the consistency of DAMA and CoGeNT. In comparison to the usual method of displaying results in the $\sigma_n$ -- $m_{\chi}$ plane, this approach has the advantage that the results do not depend on any astrophysical parameters. 
\newline

\noindent{{\bf Note added:} Following the submission of v.1 of this work to the arXiv, related work appeared: \cite{Farina:2011pw} and \cite{Fox:2011px}. Our conclusions are in agreement.

\section*{Acknowledgements}

We thank Mads T. Frandsen, Felix Kahlhoefer, John March-Russell, Matthew McCullough and Kai Schmidt-Hoberg for many stimulating discussions and the CoGeNT collaboration for making the time-stamped data publicly available. CM is supported by an STFC postgraduate studentship.

\bibliography{ref}
\bibliographystyle{ArXiv}

\end{document}